# ANTISCALAR COSMOLOGICAL BACKGROUND

Eduard G. Mychelkin[1] and Maxim A. Makukov[2]

*Fesenkov Astrophysical Institute, 050020, Almaty, Republic of Kazakhstan*
[1]edmych@gmail.com
[2]makukov@aphi.kz

**Abstract.** It is shown that the antiscalar approach to dark energy, whereby the energy-momentum tensor of the scalar field has the sign opposite to that of the rest of the matter, follows from the considerations of thermodynamic stability, as well as from the static limit of the Einstein-Maxwell equations. The same limit also demonstrates that the resulting antiscalar background field proves to be of quasi-electrostatic origin.

**Introduction**

Traditional descriptions of dark energy dynamics generally reduce to consideration of the Einstein equations which include a scalar field with a certain equation of state (EoS). The variety of existing scalar-field models of dark energy might be categorized as follows:

- Cosmological constant
- Quintessence
- K-essence
- Phantom fields
- Tahyonic fields (arising in the low-energy limit of the string theory)
- Chaplygin gas
- Dilaton field, etc.

The **cosmological constant** might be represented as perfect fluid obeying stiff vacuum EoS $p = w\varepsilon$ with $w = -1$. At that, the $\Lambda$-term in Einstein's equations

$$G_{\mu\nu} - \Lambda g_{\mu\nu} = \kappa T_{\mu\nu}, \quad \text{or} \quad G_{\mu\nu} = \frac{8\pi G}{c^4}\left(T_{\mu\nu} + \frac{\Lambda c^4 g_{\mu\nu}}{8\pi G}\right), \qquad (1)$$

is related to the energy density $\varepsilon = \varepsilon_{vac}$, with $\varepsilon_{vac} \equiv \varepsilon_{\Lambda} = \Lambda c^4 / 8\pi G$ (conventions used here are the same as in Landau & Lifshitz, 1971). As follows from (1), the dynamics of a system with an arbitrary energy-momentum tensor (EMT) for perfect fluid-like matter in isotropic coordinates $ds^2 = dt^2 - a^2(t)d\vec{r}^2$ is described by the Friedmann equation for the scale factor $a(t)$ (overdots denote derivatives with respect to time $x^0/c = t$):

$$\frac{\ddot{a}}{ac^2} - \frac{\Lambda}{3} = -\frac{4\pi G}{3c^4}(\varepsilon + 3p). \tag{2}$$

In case of the ideal de Sitter $\Lambda$-vacuum ($G_{\mu\nu} \equiv R_{\mu\nu} - Rg_{\mu\nu}/2 = \Lambda g_{\mu\nu}$, $\varepsilon = 0$, $p = 0$) with positive cosmological constant $\Lambda > 0$, equation (2) has the exponential solution ("inflation"):

$$a(t) = e^{Ht}, \quad H = \dot{a}/a = \sqrt{\Lambda c^2/3}.$$

From (2) it follows that to model the accelerated expansion of the universe one needs a medium with vacuum-like EoS $p < -\varepsilon/3$. Scalar fields represent instances for such media.

Here we briefly characterize other mentioned classes of scalar-field dark energy models.

***Quintessence*** is a standard scalar field, minimally coupled to gravity and appended with a special (*ad hoc*) self-interaction potential. This implies that the scalar sector of the theory contains (in its Lagrangian) only the kinetic term and the self-interaction potential $V(\phi)$.

In case of ***K-essence*** the role of an *ad hoc* parameter is played not by potential, but by the kinetic term itself: $X \equiv -(1/2)\nabla_\mu \phi \nabla^\mu \phi$. The Lagrangian of K-essence is a (in general, arbitrary nonlinear) function of a scalar field $\phi$ and of the mentioned kinetic term, which enters the Lagrangian in non-canonical way (e.g., quadratically).

***Phantom fields*** are characterized by the EoS parameter $w < -1$. This is achieved by means of the negative kinetic term (in the quintessence Lagrangian the kinetic term is taken now with the opposite sign). With that, certain problems related to relativistic causality arise. The *ad hoc* function here is, as before, the self-interaction potential.

***Tachyonic fields*** arising in the low-energy limit of the string theory have a factorized (rather than additive) nonlinear *ad hoc* potential and the variable EoS parameter $w_\phi = p/\varepsilon = \dot{\phi}^2 - 1$, which takes values from -1 to 0.

***Chaplygin gas*** has "inverted" EoS $p = -A/\varepsilon$ and might be considered as a special case of tachyonic field with constant potential, or as a usual minimal scalar field with a specific nonlinear potential. Chaplygin gas has a unique property: it might evolve from dust-like ($p \approx 0$) state to vacuum-like ($p \approx -\varepsilon$) state, imitating evolution of the universe (tachyonic fields might also display similar behavior).

***Dilaton fields*** arise, in particular, at dealing with quantum fluctuations of original classic fields of some of the types considered above, and they are capable then to form a condensate with the required properties, etc.

All of the listed models have EoS providing qualitatively the observed accelerated expansion of the universe, and it is difficult to decide which of them is most realistic. The common drawback of all these models is that none of them is self-consistent, as they do not provide explanations for the origin of their *ad hoc* scalar fields. In this situation, an alternative *antiscalar approach* deserves attention, since its scalar field is well defined from the outset, and its sources appear to be ordinary masses. This approach does not necessarily require an *ad hoc* potential, whereas the principle of antiscalarity, implying that the entire EMT of the scalar field enters Einstein's equations with the sign opposite to that of the rest of the matter, finds fundamental justification in thermodynamics. (The term is used in analogy with the term "anti-de Sitter", implying that the EMT of physical vacuum has the sign opposite to that of the ordinary matter).

**Relativistic thermodynamics** allows to consistently describe equilibrium states of both ordinary matter and scalar vacuum. This might be shown with the known definitions (Synge, 1957) of thermodynamic quantities as functions of some fundamental geometrical scalar $\xi$ (the modulus of a time-like Killing field), which has the meaning of the reciprocal temperature:

temperature: $T = 1/\xi$ (the Boltzmann constant is set to $k = 1$);

energy density: $\varepsilon = -\partial n / \partial \xi$;

pressure: $p = n/\xi$;

number density: $n = n(\xi)$.

Plugging these definitions into the thermodynamic Gibbs identity (with $s$ being the entropy density and $q$ the heat flux density)

$$dq = Td(s/n) = d(\varepsilon/n) + pd(1/n),$$

we find that, in the absence of heat sources ($dq = 0 \Leftrightarrow ds/s = dn/n$), this identity transforms into the following ordinary differential equation:

$$nn'' + nn'/\xi - (n')^2 = 0, \qquad n = n(\xi).$$

The first integral of this equation turns out to be nothing but the well-known barotropic EoS:

$$-w\partial n / \partial \xi = n/\xi \Leftrightarrow p = w\varepsilon,$$

where $w$ represents the constant of integration (often denoted also as $\gamma - 1$). Integrating once more, we obtain a relation with another constant $C = n_0 \xi_0^{1/w}$:

$$n = C\xi^{-1/w}.$$

Denoting $z = 1 + 1/w$, we represent all thermodynamic quantities as functions of $\xi$ (here $s$ is the entropy density at zero chemical potential, $\mu = 0$):

$$\varepsilon = -\partial n / \partial \xi = (C/w)\xi^{-z},$$

$$p = n/\xi = C\xi^{-z},$$

$$s = dp/dT = \xi(\varepsilon + p) = Cz\xi^{-1/w}.$$

In general case, $s = \xi(\varepsilon + p - \mu n)$, where $\varepsilon$, $p$ and $n$ retain their previous dependence on $\xi$, and so finally we have:

$$s = C(z - \mu)\xi^{-1/w}.$$

For $w = 1$ (typical stiff state of a scalar field) we obtain the basic relations for the scalar vacuum thermodynamics:

$$\varepsilon = -\partial n / \partial \xi = C\xi^{-2}, \qquad p = n/\xi = C\xi^{-2}, \qquad s(\mu = 0) = dp/dT = \xi(\varepsilon + p) = 2C\xi^{-1}.$$

**Criterion of thermodynamic stability of a scalar field**

Substituting the latter relations into the thermodynamic identity $p = n\partial\varepsilon/\partial n - \varepsilon$, we find that the criterion for thermodynamic stability $\partial^2\varepsilon/\partial n^2 > 0$ assumes the following form:

$$w(w+1)n^{w-1} > 0.$$

In particular, this criterion is satisfied for $w = 1$, implying that a typical scalar field represents a stable medium.

**Thermodynamic justification of antiscalar approach**

Contracting the Einstein equations with perfect fluid $p = w\varepsilon$ and with positive cosmological constant, and taking thermodynamic relations obtained above into account, we get:

$$G_{\mu\nu} = 8\pi G \left(T_{\mu\nu}\right)^{perf} + \Lambda g_{\mu\nu} \quad \Rightarrow \quad 8\pi G C(1/w - 3)T^{1+1/w} = -R - 4\Lambda, \qquad (3)$$

where $-R > 0$. For stiff EoS with $w = 1$ expression (3) yields negative square of temperature, i.e., a meaningless result. To avoid that, it is necessary (and sufficient) to change the sign of the EMT of the scalar field. The sufficiency follows from the fact that only negative EMT of scalar field leads to correct solutions consistent with observations.

Being an inherent characteristic of the cosmological background, this negative (antiscalar) EMT cannot be removed from the field equations whatsoever. Thus, vacuum Einstein's equations and their solutions (black holes, gravitational waves) cannot be considered physically adequate from the standpoint of antiscalarity.

In general case, the original EMT, by itself, should include a constant component, or the customary (positive) cosmological term, as well as the mass-term, both of which switch the sign to the opposite due to antiscalarity:

$$\left(T_{\mu\nu}\right)^{scalar} = \phi_\mu\phi_\nu - \tfrac{1}{2}g_{\mu\nu}\left(\phi_\alpha\phi^\alpha - \mu^2\phi^2\right) + (\Lambda/8\pi G)g_{\mu\nu},$$

where, for quasi-static scalar field which imitates instant action, we adopt $\mu^2 = -m^2$. Besides, under the integrability condition of the corresponding Einstein equation (de Siqueira, 2001) $\Lambda = |\Lambda| = -2m^2/3$, it follows:

$$ds^2 = dt^2 - e^{-|\Lambda|(t-t_0)^2}(dr^2 + r^2 d\Omega^2). \tag{4}$$

Thus, thermodynamics requires the following representation of the Einstein equations:

$$G_{\mu\nu} = -8\pi G (T_{\mu\nu})^{scalar} + 8\pi G (T_{\mu\nu})^{matter}.$$

Then, for all stiff media with EoS in allowed limits $1/3 < w \leq 1$ and with the cosmological constant taken into account, we have from (3):

$$8\pi G C (3 - 1/w) T^{1+1/w} = -R + 4\Lambda, \quad (-R > 0).$$

In particular, for $w = 1$ one obtains sensible relation:

$$16\pi G C T^2 = -R + 4\Lambda > 0, \quad (-R > 0).$$

It is worth noting that in the limit of large scales, from here one finds relations coinciding (up to a constant factor) with quasi-classic Hawking formulas for the de Sitter vacuum:

$$T \propto const\sqrt{\Lambda}, \quad s \propto const\sqrt{\Lambda}.$$

However, this time they represent universal asymptotic relations obtained from relativistic thermodynamics independently of the specific form of metric and expressing the connection between temperature and entropy of scalar background with cosmological constant.

So, antiscalar approach to dark energy is an alternative to the existing models and appears to be physically adequate, since it has fundamental thermodynamic justification, as well as serious support from electrodynamics in curved space-time, to which we now turn.

**The static limit of the Einstein-Maxwell equations and the cosmic vacuum**

It turns out that the antiscalarity principle also follows from the static limit of the Einstein-Maxwell equations, showing that the background scalar field has electrostatic nature.

Indeed, in *static limit*, when electromagnetic field reduces to the scalar Coulomb field, the Einstein-Maxwell equations take the same form as equations of *antiscalar gravitation with massless scalar field* (it is clear that scalar fields considered here might have only extremely small masses, which might be neglected at least in local experiments). Namely, the sign of the EMT of the Coulomb field is opposite to the sign of the EMT of ordinary matter:

$$G_{\mu\nu} = 8\pi G (T_{\mu\nu})^{elm} \quad \Rightarrow \quad G_{\mu\nu} = -8\pi G (T_{\mu\nu})^{scalar},$$

where the massless EMT of the Coulomb field has the standard form:

$$\left(T_{\mu\nu}\right)^{scalar} = \phi_\mu \phi_\nu - \tfrac{1}{2} g_{\mu\nu} \phi_\alpha \phi^\alpha.$$

Full correspondence between antiscalar approach and electrostatics becomes especially obvious if one considers the gravielectric balance. Thus, in case of two or more gravitating (like) charges, their electric repulsion is exactly cancelled out by gravitational attraction, provided that:

$$e_i = \pm\sqrt{G}m_i, \quad \text{or, in units } G=c=1, \quad e_i = \pm m_i,$$

and similarly for mass and charge densities (Majumdar, 1947; Papapetrou, 1947; Das, 1962; Bonnor, 1981; Mann & Ohta, 2000). Under this condition (accurate to the mentioned dimensional constants), solutions of the equations of antiscalar approach and of the static Einstein-Maxwell equations coincide with the Papapetrou solution (Papapetrou, 1954):

$$ds^2 = e^{-2m/r} dt^2 - e^{2m/r}\left(dr^2 + r^2 d\Omega^2\right). \tag{5}$$

This coincidence (balance) of electric $e_i$ and scalar (often called gravitational) $\sqrt{G}m_i$ charges suggests that the quasi-static Coulomb and scalar fields might have common origin in a sense that neutral superposition of electric fields generated by fermionic matter looks like a gravitating scalar field:

$$(\phi_+ + \phi_-)/\sqrt{2} = \phi.$$

In this regard, masses might be considered as special states and sources of the same field.

An arbitrarily small imbalance between $\phi_+$ and $\phi_-$ on any scales manifests itself as the electric field. With that, electromagnetic interaction of real fermions is many orders of magnitude stronger than gravitational one (e.g., for an electron and a proton, by a factor of $\sim 10^{39}$).

**Masses are scalar charges, scalar charges are masses**

Historically, the notion of a universal scalar field, developed by Papapetrou (1947; 1954), had arisen exactly in considering electric fields on the basis of the Einstein-Maxwell equations. However, the correspondence between the sources of (anti)scalar field (scalar charges) and masses of gravitating bodies might be demonstrated also independently of the Einstein-Maxwell equations.

To this end, consider the limit $b \to 0$ in the Janis-Newman-Winicour spherically-symmetric solution of Einstein's equations with minimal scalar field (Janis et al., 1968), whose source possesses the scalar charge $q$ (Virbhadra et al., 1998):

$$ds^2 = \left(1-\frac{b}{r}\right)^\gamma dt^2 - \left(1-\frac{b}{r}\right)^{-\gamma} dr^2 - \left(1-\frac{b}{r}\right)^{1-\gamma} r^2 (d\theta^2 + \sin^2\theta d\varphi^2).$$

The solution of the corresponding Klein-Gordon equation is:

$$\phi = \frac{q}{b} \ln\left(1 - \frac{b}{r}\right).$$

Here $\gamma = 2m/b$, $b = 2\sqrt{m^2 + q^2}$, and we adopt here $G = c = 1$. Obviously, at $b \to 0$, from here the fundamental Papapetrou solution (5) follows, for which, as it is known, all "crucial effects" hold. At the same time, the asymptote $b \to 0$ implies

$$q^2 = -m^2 \leftrightarrow q = im,$$

i.e., the equality of the absolute value of the scalar charge $q$ and of the source mass. With that, the resulting scalar field is also multiplied by the imaginary unit:

$$\phi = \frac{q}{b}\ln\left(1 - \frac{b}{r}\right) \to q/r = im/r = i\bar{\phi}, \quad \text{i.e. } \phi \to i\bar{\phi}, \qquad (6)$$

which is necessary for the consistency of the limit $b \to 0$. Then in the original Einstein equations one should change the sign of the scalar field EMT (quadratic in $\phi$) to the opposite. But this implies transition to antiscalar theory dealing with the Papapetrou metric (5), and thereby justified the appearance of the imaginary unit in (6).

Thus, on the one hand, analysis of the static limit of the Einstein-Maxwell equations reveals that background scalar field might be represented as a superposition of electric fields related to the fermionic matter in the universe. On the other hand, in antiscalar approach masses are the sources ("charges") of the universal scalar field which is responsible for all gravitational effects. The known quasi-classic effects (Hawking effects) related to black hole thermodynamics hold true qualitatively for correspondingly small compact objects with characteristic sizes of the order of their gravitational radius.

**Conclusion**

The antiscalar approach discussed here (the negative EMT of massive tachyonic scalar field with $\Lambda$-term) is justified thermodynamically and geometrically, and dismantles some of the problems. Thus, in this approach the problem of the "gravitational field" energy does not arise, since the task now can be reduced to finding the energy of gravitating (anti)scalar field, which is well-defined. E.g., for a single gravitating mass (e.g., a star or a galaxy) the total energy of the scalar field is positive and is equal to $Mc^2$. The absence of the solutions of the type "gravitational waves propagating at the speed of light" erases the problem of their quantization. As for the background (anti)scalar field, it also cannot be canonically quantized due to its tachyonic nature: this is the classical subquantum vacuum. At the same time, all "crucial experiments" hold, and quantum effects in external (classical) fields retain their meaning. Unlike other approaches, in our case the scalar field has a definite origin: as follows from the comparison with the Einstein-Maxwell equations, it is generated by all fermionic matter and represents the neutral superposition of quasi-static electric fields: $\phi \sim \phi_+ + \phi_-$.

Next, following Schwinger (1973), we admit that quasi-static fields have their own carriers ("statons"), whose tachyonic mass, as follows from the conditions of integrability of the field equations, is extremely small but finite: $|\Lambda| = -\frac{2}{3}m^2$, $|m| = m_\phi \approx 10^{-33} eV \approx 10^{-65} g$. With that, the known fundamental de Sitter solution is replaced with the more realistic metric (4), which turns any "inflationary" process back to its original state. This is a requisite for the oscillating model of the universe.

The "electrical" nature of the cosmological scalar field established here implies the existence of a similarly tiny but finite mass of the hypothetical tachyonic carriers $m_{\phi_\pm} \sim 10^{-33}$ eV for the Coulomb fields as well (Schwinger, 1973). Strictly speaking, this leads to inevitable modification (without any effects measurable in modern experiments) of the equations of Maxwellian electrodynamics, and, in principle, after extending this approach to non-Abelian fields – of the entire Standard Model. This topic is to be discussed elsewhere.